\documentclass[preprint2]{aastex}

\newcommand{\sax}{{\it BeppoSAX\,\,}}
\newcommand{\lae}{\mathrel{<\kern-1.0em\lower0.9ex\hbox{$\sim$}}}
\newcommand{\gae}{\mathrel{>\kern-1.0em\lower0.9ex\hbox{$\sim$}}}
\newcommand{\ecs}{{\rm erg} \, {\rm cm}^{-2} \, {\rm s}^{-1}} 
\newcommand{\cps}{\,{\rm counts \, s}^{-1} }
\newcommand{\keV}{\,{\rm keV} }
\newcommand{\mJy}{\,{\rm mJy}}

\newcommand{\GHz}{\,{\rm GHz}}

\shorttitle{Gamma-ray bright BL Lac object RX J1211+2242}
\shortauthors{Beckmann et al.}

\received{2004 January 6}

\begin{document}

\title{The Gamma-ray bright BL Lac object RX J1211+2242}


\author{V. Beckmann\altaffilmark{1}}
\affil{NASA Goddard Space Flight Center, Code 661, Greenbelt, MD 20771, USA}
\email{beckmann@milkyway.gsfc.nasa.gov}
\author{P. Favre\altaffilmark{2}}
\affil{INTEGRAL Science Data Centre, Chemin d' \'Ecogia 16, CH-1290
 Versoix, Switzerland}
\author{F. Tavecchio}
\affil{Osservatorio Astronomico di Brera, via E. Bianchi 46, I-23807 Merate, Italy}
\author{T. Bussien}
\affil{\'Ecole Polytechnique F\'ed\'erale de Lausanne, CH-1015 Lausanne, Switzerland}
\author{J. Fliri}
\affil{Universit\"ats-Sternwarte M\"unchen, Scheinerstra{\ss}e 1, D-81679 M\"unchen, Germany}
\and
\author{A. Wolter}
\affil{Osservatorio Astronomico di Brera, via Brera 28, I-20121 Milano, Italy}


\altaffiltext{1}{Joint Center for Astrophysics, Department of Physics, University of Maryland, Baltimore County, MD 21250, USA}
\altaffiltext{2}{Observatoire de Gen\`eve, 51 Ch. des Maillettes, CH-1290 Sauverny, Switzerland}

\begin{abstract}
RX J1211+2242 is an optically faint 
($B \simeq 19.2 \, \rm mag$) but X-ray bright ($f_{2 - 10 \keV} = 5 \times 10^{-12} \ecs$) AGN, which has been shown to be a BL Lac object at redshift $z = 0.455$. The {\it ROSAT} X-ray, Calar Alto optical, and NVSS radio data suggest that the peak of the synchrotron emission of this object is at energies as high as several keV. {\it BeppoSAX} observations have been carried out simultaneously with optical observations in order to extend the coverage to higher energies. The new data indeed indicate a turn-over in the $2 - 10 \keV$ energy region.
We propose that RX J1211+2242 is the counterpart of the unidentified EGRET source 3EG J1212+2304, making it a gamma-ray emitter with properties similar to, for example, Markarian 501 in its bright state, though being at a much larger distance. 
\end{abstract}


\keywords{BL Lacertae objects: general --- BL Lacertae objects: individual (RX J1211+2242) --- gamma rays: observations}


\section{Introduction}
The nature of astronomical sources in the gamma-ray domain above 20 MeV 
is still not well understood. While we have identified most of the sources seen at energies
below 1 MeV, the identification of sources at higher energies suffers from large uncertainties in their positions and in many cases strong variability. The Energetic Gamma-Ray Experiment Telescope (EGRET, Kanbach et al. 1988) on board the {\it Compton Gamma Ray Observatory} ({\it CGRO}) covered the energy range between 20 MeV to over 30 GeV. But, even ten years after the mission, most of the 271 sources listed in the 3rd EGRET catalog are still unidentified \cite{hartman}. 
While many sources in the Galactic plane are associated with populations of massive stars \citep{kaaret,yadi}, the largest fraction of high Galactic latitude sources turned out to be AGNs. Among this latter group, blazars are dominant.

The common model 
of the blazar phenomenon is that we are looking down 
a highly relativistic jet when observing these sources. This leads to the characteristic properties of blazars, such as high polarization, high variability, and, for some objects, featureless optical spectra. The overall spectral energy distribution (SED) can be described by two components \cite{pado}. The 
first component is thought to be due to synchrotron radiation from relativistic electrons (in some models also positrons) interacting with the magnetic field of the jet.  The peak emission of this component occurs in the infrared region for Flat Spectrum Radio Quasars, in the UV range for low frequency cut-off BL Lacs (LBL), and mostly in the X-ray domain for high frequency cut-off BL Lacs (HBL). 
The second component can be well modeled by inverse Compton scattering of photons by relativistic electrons. Here the seed photons can be either the synchrotron photons themselves (SSC; e.g. Mastichiadis \& Kirk 1997) or external photons produced by the accretion disk \cite{EC} and/or the broad line region \cite{sikora}. Due to relativistic beaming towards the observer, the emission of blazars is amplified, which makes them the brightest extragalactic objects, especially around the observed peak frequencies. This also explains their dominance in the EGRET energy band, where the peak of the inverse Compton emission 
is expected. 

Here we present the blazar RX J1211+2242 and show that it is likely to be associated with the unidentified EGRET source 3EG J1212+2304.

\section{First observations of RX J1211+2242}
The X-ray source RX J1211+2242 was discovered in the course of the {\it ROSAT} All-Sky Survey (RASS; Voges et al. 1999) with a flux of $f_{0.5 - 2.0 \keV} = 2.6 \times 10^{-12} \ecs$. 
The position of the X-ray emission coincides with the radio source NVSS J121158+224232, detected in the NRAO VLA Sky Survey (NVSS; Condon et al. 1998), with a $1.4 \GHz$ flux of $19.7 \mJy$. The distance between the RASS and the NVSS 
position is $8.9''$, well inside the RASS error circle of $\sim 30''$.
Due to its high X-ray flux and its detection in the radio domain, RX J1211+2242 was selected as a candidate for the Hamburg/RASS X-ray selected BL Lac sample (Beckmann et al. 2003). An optical spectrum to identify this X-ray source was taken in February 1998 with the Calar Alto 3.5m telescope, confirming its BL Lac nature and revealing absorption features due to the host galaxy, which gave a redshift of $z = 0.455$ \cite{hrxbllac}.
Observations in the near infrared taken during the Two-Micron All-Sky Survey (2MASS; Skrutskie et al. 1995) yielded  magnitudes of $J = (17.38 \pm 0.25) \, \rm mag$, $H = (16.33 \pm 0.21) \, \rm mag$ and $K = (15.52 \pm 0.15) \, \rm mag$. 
The available data indicate that the source might have the peak of its synchrotron emission in the X-ray region \cite{phd}.
To confirm this, simultaneous observations in the optical at Calar Alto 1.23m telescope and in the X-rays  using the {\it BeppoSAX} satellite were carried out from December 2001 to January 2002. 

\section{Observations}
\subsection{BeppoSAX}
The X-ray observations were made with {\it BeppoSAX}, a project of
the Italian Space Agency with participation of the Netherlands
Agency for Aerospace. A detailed description of the \sax
mission can be found in Butler \& Scarsi (1990) and Boella et al. (1997a). The data presented here are from the Low Energy
Concentrator Spectrometer (LECS)
and the Medium Energy Concentrator Spectrometer (MECS).
The Phoswich Detector System (PDS)
did not detect the source.

The data have been pre--processed at the \sax Science Data
Center (SDC) and retrieved through the SDC archive.  Table~\ref{journal}
shows the journal of observations, including exposure times and net
count rates for the LECS and MECS detectors. 




The spectra from the MECS units were summed together to increase
the S/N. 
The reduction process described in Wolter et al. (1998) has been applied to 
this data in the energy ranges $0.12 - 4.0 \keV$ (LECS) and $1.65 - 10.5 \keV$
(MECS), using FTOOLS 4.0 and XSPEC 11.0.

\begin{figure}
\plotone{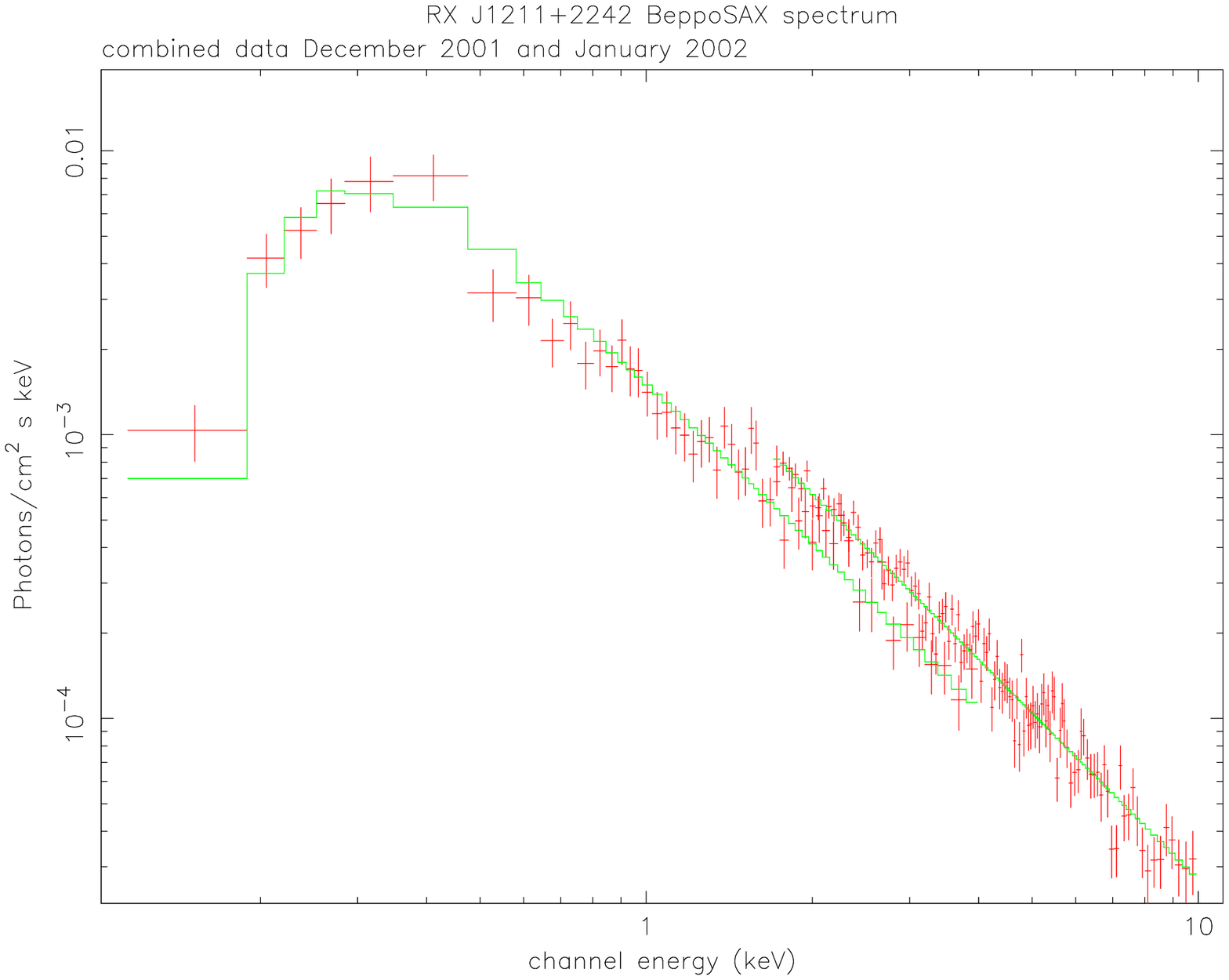}
\caption[]{\sax spectrum of RX J1211+2242. Combined data from the December 2001/January 2002 observations. The model shows a single power law with $\alpha_X = 0.95$ and $N_H = 2.9 \times 10^{20} cm^{-2}$.}
\label{fig:combined}
\end{figure}

We simultaneously fit the LECS and MECS data, leaving as a free parameter the LECS
normalization with respect to the MECS to account for the residual
errors in the intensity cross-calibration. A reliable value
for the ratio should be $\rm LECS/MECS \sim 0.7 \pm 0.1$.
The data were fit using a power law spectral model with energy index $\alpha_X$ modified by interstellar photoelectric absorption, using
the photoelectric absorption cross sections of Morrison \& McCammon (1983).
The Galactic hydrogen column density, $N_{H,gal}$, as derived by the Leiden/Dwingeloo Survey \cite{LDS} is $2.52 \times 10^{20} \, \rm cm^{-2}$.
We list the best fit parameters in Table~\ref{powfreesax}.
The errors are 90\% confidence levels, simultaneously determined for the three interesting parameters ($N_H$, $\alpha_X$, and LECS/MECS normalization). Fluxes in the 2--10 keV and 0.5--2.0 keV bands (for comparison with
the ROSAT-PSPC flux) are given. In addition, the normalization
factors of the LECS relative to the MECS are listed. 
The December 2001 and January 2002 data are two subsets of the same observation.  The data were taken within a three week time span and can be combined since the source did not show significant variations during that time.
In order to test the variability, the data have been binned into 5000 sec bins spanning the combined data set. Under the assumption that the flux is constant, a $\chi^2$ test gives $\chi^2/dof = 1.06$ and $\chi^2/dof = 1.09$ in the low and high energy bands, respectively. The MECS 2--5 keV mean source count rate is $(6.338 \pm 0.009) \times 10^{-2} \cps$, and the 5--10 keV mean rate is $(2.452 \pm 0.002) \times 10^{-2} \cps$. 
In the following we have therefore used the combined December 2001 and January 2002 data set.  
Fitting the combined data results in the photon spectrum with $\alpha_X = 0.95$ shown in Fig.~\ref{fig:combined}.  
The December 1999 observation has a similar slope ($\alpha_X = 1.08$) to this combined spectrum with a flux lower by a factor 2.9, which is not surprising since BL Lacs are known to be variable sources \cite{WaWi}.   

All observations show a small excess with respect to the Galactic value in the range $0.4 - 1.0 \times  10^{20} \, \rm cm^{-2}$, which is  consistent at a $1\sigma$ level when compared to the Galactic value. The spectral slope is always consistent with a flat spectrum in the $\log \nu f_\nu - \log \nu$ diagram. Thus the energy range covered by the \sax LECS and MECS detectors has to be close to or even contain the peak of the synchrotron emission, where a change in slope, from a rising (i.e. $\alpha_X < 1.0$) to a falling (i.e. $\alpha_X > 1.0$) spectrum is expected.  
To try a more complex model we applied a broken power law. This shows that the peak of the spectrum occurs within the MECS energy band at $2.3 {+2.0 \atop -0.6} \rm \, keV$ with an energy index of $0.82 {+0.05 \atop -0.10}$ and $1.02 {+0.14 \atop -0.06}$ below and above the break energy, respectively.
However, the complex spectrum is not required statistically since, according to the F-test, the broken power law does not improve the fit compared to the single power law model.

\subsection{Optical observations}
BL Lac objects are also known to be variable sources in the optical band \cite{WaWi}. 
Even though X-ray dominated BL Lacs with higher cut-off energies have been shown to be less variable in the optical range than those with low cut-off energies (see e.g. Heidt \& Wagner 1998, Villata et al. 2000), simultaneous data are still needed. 
Optical observations have been carried out using the 1.23m telescope at the Calar Alto observatory. Absolute photometry was obtained in April 2000 in order to calibrate stars in the field of RX J1211+2242 \cite{phd}. Four observations simultaneous with the \sax exposure were performed during two nights, December 27 and 28, 2001. The list of optical observations is given in Table \ref{journalobs}, labeling the two observations per night with A and B.
In the following, we use the mean value of 
$B = 19.19 \, \rm mag$. The small difference between this and the April 2000 measurement ($B = 19.08 \, \rm mag$) is consistent with the finding of Ulrich et al. (1997). They showed that variability is larger at or above the peak  of the synchrotron emission (in our case in the X-rays) than below (i.e. in the optical domain).

\section{The unidentified EGRET source 3EG J1212+2304}
\begin{figure}
\plotone{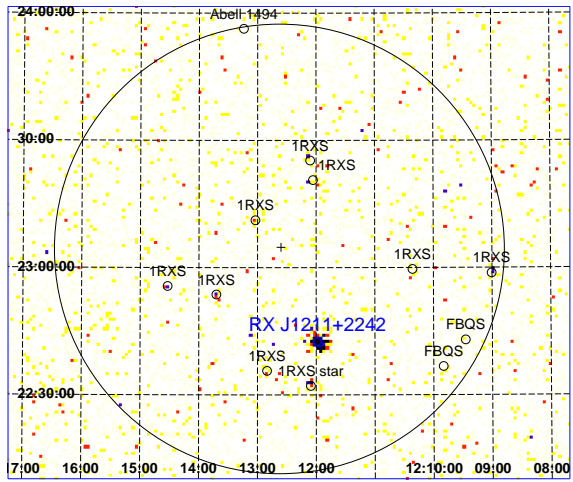}
\caption[]{{\it ROSAT} All-Sky Survey map ($0.1 - 2.4 \keV$) around the position 3EG J1212+2304. The circle marks the 53 arcmin EGRET error radius. There is only one bright X-ray source in the error circle, which is the blazar RX J1211+2242}
\label{fig:RASSchart}
\end{figure}

Most of the 
point sources detected by EGRET are still unidentified \cite{hartman}.
We found 271 entries in the NED and 156 in the SIMBAD archive within the 53' $95 \%$ confidence radius of 3EG J1212+2304.
Since the objects in those databases overlap, we merged the results. Most of the entries are either normal galaxies or stars, which have not been detected at any wavelength other than optical and which are not 
expected to be gamma-ray emitters. Of the rest, 77 are from infrared and radio catalogs (e.g. from the NVSS). Two are quasars from the FIRST Bright Quasar Survey \cite{FBQS} at redshift 1.722 and 1.208. Those have not been detected by the RASS and are therefore less likely to be the counterpart.
In total there are only ten X-ray sources detected in this region in the {\it ROSAT} All-Sky Survey, and only RXJ 1211+2242 is both in the {\it ROSAT} Bright Source Catalog and 
the NVSS radio catalog. 
Figure~\ref{fig:RASSchart} shows a map derived from the RASS in the $0.1 - 2.4 \keV$ energy range, centered on the position of the gamma-ray source 3EG J1212+2304. The positions of the other {\it ROSAT} sources are marked with ``1RXS'' and the positions of the two quasars and an Abell galaxy cluster inside the error circle are also indicated.

The gamma-ray source shows variability. Its position was covered by 22 EGRET observations, in three of which it was clearly detected at different flux levels.
The photon flux measurements vary from the lowest upper limit of $5.3 \times 10^{-8} \, \rm photons \, cm^{-2} \, s^{-1}$ to a detected flux of $(50.8 \pm 16.6) \times 10^{-8} \, \rm photons \, cm^{-2} \, s^{-1}$. Applying the conversion as described in Thompson et al. (1996) and assuming a photon index of $\Gamma = 2$ in the EGRET energy band, this corresponds to a flux range of $f_{0.1-5 \rm \, GeV} = (<3.4 \dots 32.5) \times 10^{-11} \ecs$. The light curve, covering a time span of 4 years, is displayed in Fig.~\ref{fig:EGRETlc}. Most data points represent an integration time of one week. The detection in 1992/1993 refers to the analysis of the whole {\it CGRO} cycle 2 EGRET data, which lasted from 1992 November 17 until 1993 September 7. 
The difference in the upper limit values depends on the variable sensitivity of the measurement, due to different positions of the source in the EGRET FOV, different exposure times, etc. For details on the EGRET data see Hartman et al. (1999) and Esposito et al. (1999).
The fact that the source has to be a highly variable object in the gamma-ray domain, together with the measurements at lower frequencies, plus 
the lack of another plausible candidate inside the error circle except the BL Lac object, make it probable that  RXJ 1211+2242 is indeed the counterpart of 3EG J1212+2304.

\section{Spectral energy distribution}
\begin{figure}
\plotone{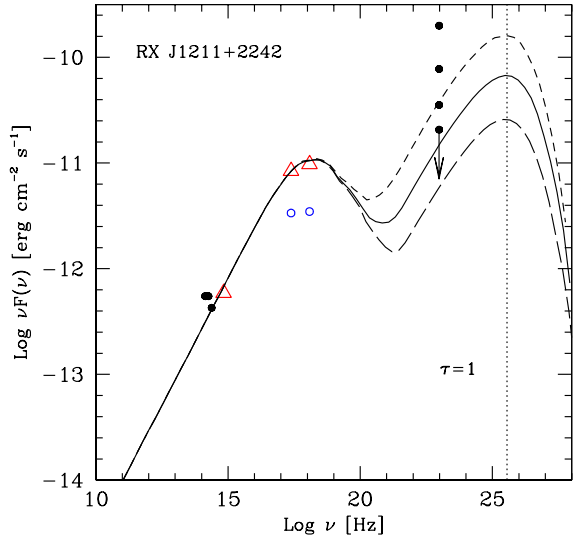}
\caption[]{Spectral energy distribution of RX J1211+2242 constructed assembling the available data. The triangles mark the simultaneous optical and X-ray data ($0.5 - 2.0 \rm \, keV$ and $2 - 10 \rm \, keV$ bands). The curves
represent the SED calculated with the model described in the text,
with the following parameters for all the curves: $B=0.1$ G, $n1=2$,
$n2=4$, $R=5\times 10^{15}$ cm. The three models differ for the value
of $\delta =10, 15, 20$ (short dashed, solid, long dashes,
respectively), the normalization of the electron density $K=8, 2.3,
1\times 10^5$ part/cm$^{3}$ and the break Lorentz factor $\gamma _b=6,
4.5, 3\times 10^{5}$. The frequency where absorption by infrared photons becomes important \cite{kneiske} is indicated by the dotted vertical line.}
\label{fig:SED}
\end{figure}

Combining the data from the NVSS radio catalog and the 2MASS with the simultaneous optical and \sax data, we can reconstruct the synchrotron branch of the SED. We also include in the SED, which is shown in Fig.~\ref{fig:SED}, the 3 EGRET detection data points and the lowest upper limit, following the assumption that 3EG J1212+2304 is the gamma-ray counterpart of RX J1211+2242. 

Various authors have applied a parabolic fit to the synchrotron branch of the SED of BL Lacs from various samples (e.g. Fossati et al. 1998) and showed that this approximation is valid.
We have thus applied a parabolic fit in the $\log \nu f_\nu - \log \nu$ plane -- parameterized as $\log \nu f_\nu = a \cdot (\log \nu)^2 + b \cdot \log \nu + c$ -- and determined the peak frequency ($\nu_{peak}$) of the synchrotron component of RX J1211+2242. Using only the simultaneously observed data, the peak of the synchrotron component appears to be at $6 \keV$.
Using both simultaneous and non-simultaneous data yields a curve which would peak at frequencies higher than physically possible (with a peak of the synchrotron branch at $\nu_s > 10^{22} \rm \, Hz$) 
and which also does not reproduce the flat ($\alpha_X = 0.95$) X-ray spectrum.
We have also tried to fit the radio, infrared, and optical data together with the December 1999 \sax observation, and this gives again a peak in the X-ray domain, around 22 keV.

Although the data do not completely cover the SED of RX J1211+2242 one can draw from this model a few conclusions about the inverse Compton (IC) emission of this BL Lac.
A simple approach is the one proposed by Stecker et al. (1996) who assumed that the SED of HBLs is in general similar to those of Markarian 421 and PKS 2155-304. In these objects, the ratio of the peak frequencies of the IC, $\nu_c$, and synchrotron branch, $\nu_s$, are $\nu_c / \nu_s = 10^9$.  For RX J1211+2242 this would lead to an IC cut-off frequency of $\nu_c \sim 10^{27} \, \rm Hz$, in the TeV region. The ratio of the luminosities of the synchrotron and IC branches in this model is $L_c/L_s \sim 1$.

We also derive a more elaborate description of the SED using the emission model described e.g. in Maraschi et al. (1999). In this model a spherical emitting region is filled with relativistic
electrons of density $K$ with a broken-power law distribution with indices $n_1$ and
$n_2$ below and above the break energy, $mc^2\gamma _b$.  This 
region is moving with bulk Lorentz factor $\Gamma $ at an angle
$\theta $ with respect to the line of sight. Relativistic effects are
completely described by the relativistic Doppler factor $\delta
=\left[ \Gamma (1-\beta \cos \theta) \right]^{-1}$. 
As discussed in Tavecchio et al. (1998), the determination of
the physical parameters in the synchrotron/SSC model requires
knowledge of the position and luminosity of both the synchrotron
and SSC peak. Since this is not possible in the case discussed
here (for which we have simultaneous information for the optical and
X-rays only), we report possible fits of the SED assuming ``typical'' values of
the physical quantities for this kind of sources (e.g. Ghisellini
et al. 1998; Maraschi \& Tavecchio 2003), namely $\delta $ in the
range 10-20, magnetic field $B\simeq 0.1$ G, and size $R\sim 5\times
10^{15} \, \rm cm$.  We also
assume that the synchrotron peak lies in the {\it BeppoSAX} band, around 5 keV. 
Different values of $\delta$, $K$,
and $\gamma _b$ are used to match the lowest EGRET detection, the lowest upper limit and a lower flux level,
respectively, with the high-energy Compton component.
Figure \ref{fig:SED} shows the SED for these different models.
The model parameters are given in the caption.

Since RX J1211+2242 is located at a relatively large redshift, the GeV-TeV
radiation suffers severe absorption by the optical-IR background. The
optical depth for $\gamma-\gamma$ absorption has recently been
calculated by Kneiske et al.~(2003), extending previous works on this subject (e.g. de Jager \& Stecker 2002) to larger redshift. Their results have been taken into account when modeling the SED of RX J1211+2242. It appears that, for a source located at a redshift of $z\simeq 0.4$, the photons above $\sim 200
\, \rm GeV$ (corresponding to the dotted vertical line in Fig.~\ref{fig:SED}) start to be severely absorbed
before reaching the observer, thus making it impossible to be 
detected  by most of the current TeV Cherenkov telescopes which have a
high energy threshold.

\section{Discussion and Conclusion}
The RX J1211+2242 data presented here, together with previous observations\\ \cite{hrxbllac} indicate that this source is a BL Lac object with a high synchrotron peak frequency in the $2 - 10 \keV$ energy range. The object exhibits strong variations in the X-ray domain on long term time scales (years), though no significant variability was detected during the individual \sax observations.

\begin{figure}
\plotone{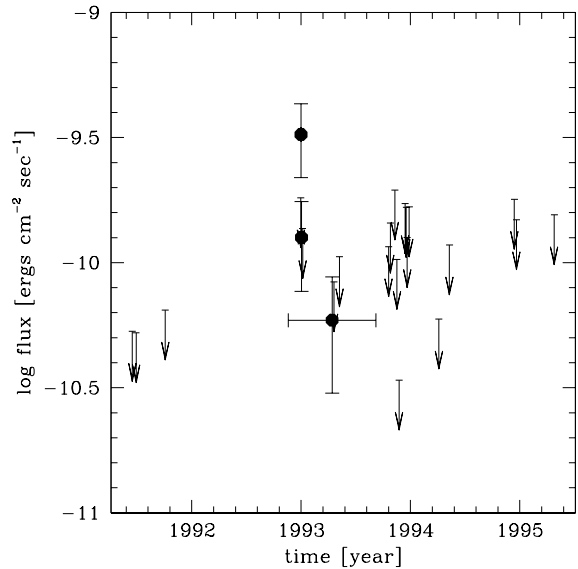}
\caption[]{EGRET lightcurve of 3EG J1212+2304 between 1991 and 1995 in the $0.1 - 5 \rm \, GeV$ energy band. Values without error bars mark upper limits or marginal detections.}
\label{fig:EGRETlc}
\end{figure}

RX J1211+2242 is located inside the 95\% confidence radius of 3EG J1212+2304,
which has been shown to be a highly variable source (see Fig.~\ref{fig:EGRETlc}). Galactic 
sources, such as pulsars and supernova remnants show little variability in the EGRET observations, while AGN exhibit strong flux variations \cite{variabilityEGRET}. 
This is a first indication that 3EG J1212+2304 might be an AGN. 
At a Galactic latitude of $b = 80.02^\circ$, this source is far out of the Galactic plane, making an extragalactic origin more likely, and thus
supporting the blazar identification.
RX J1211+2242 is the only plausible counterpart to this source inside the 
error circle.
Sowards-Emmerd et al. (2003) claim that 3EG J1212+2304 is not a blazar but base this on the evidence that there is no radio flux enhancement towards the center of the (large) EGRET error circle and that there 
was no known blazar inside the error circle, which we have just shown is not the case. 

Fitting the spectral energy distribution to a Synchrotron Self Compton model shows that physically meaningful parameters can explain the overall data, including the EGRET measurements. 
Although BL Lacs are known to exhibit a strong variability in the inverse Compton branch, it is unlikely that the {\it BeppoSAX} observations detect the source in an outburst phase due to an expected duty cycle $\ll 0.5$.
On the other hand, most of the measurements of 3EG J1212+2304 are upper limits and the modeled IC branch falls well within the region of possible flux variations. Furthermore, no other X-ray source inside the error circle is likely to be the counterpart, and no strong radio source has been found there either. Since all the identified EGRET blazars have also been detected in the RASS, a connection between a strong {\it ROSAT} source and an EGRET detection is even more likely.

Assuming that the BL Lac and the gamma-ray emitter are in fact the same source, this yields a SED similar to that of Markarian 501 (z=0.034) in its bright state. 
An emitting region of $R = 5 \times 10^{15} \rm \, cm$, a magnetic field of $B = 0.1 \rm \, G$, Doppler factor of $\delta = 20$ and a Lorentz factor for electrons at the break $\gamma_b = 3 \times 10^4$ reproduce the SED of RX J1211+2242 under the assumption that the flux during the {\it BeppoSAX} and optical observation was below the EGRET upper-limit of 3EG J1212+2304. 
The SED would reach the EGRET detections by using $\delta = 10$ and $\gamma_b = 6 \times 10^4$. In order to match the different detections and non-detections, the IC branch flux needs to vary by 
a factor of $\sim 50$ at the peak of the emission.
Again, this behavior is similar to that observed in Markarian 501 that showed variability in the hard X-rays (10-100 keV) by a factor of $\sim 100$ in different {\it BeppoSAX} observations coupled with a simultaneous shift of the synchrotron peak from the soft X-rays up to the soft gamma-rays. It also showed a strong variability, by a factor of $\sim 100$, in the EGRET band \cite{mrk501var}. 

The luminosity of the inverse Compton branch seems to be much larger than the synchrotron luminosity, at least during those periods in which the source was detected by EGRET.
If the association of the EGRET source with the BL Lac is correct, 
this would make RX J1211+2242 a peculiar 
HBL, which usually show $L_c/L_s \sim 1$ (e.g. Fossati et al. 1998). A value of $L_c/L_s \gg 1$ has already been seen in some HBLs (e.g. bright state of Markarian 501 and Markarian 421), and 1426+428 has shown an even larger 
dominance of the IC branch ($f_{IC} > 10^{-10} \ecs$; Aharonian et al. 2003) over the synchrotron component ($f_s \simeq 2 \times 10^{-11}  \ecs$; Costamante et al. 2001). It also must be taken into account that the bright gamma-ray state of this object is likely to be accompanied by increased synchrotron emission, which would decrease the $L_c/L_s$ ratio to values previously
observed in HBL. With respect to the blazar sequence \cite{fossati}, RX J1211+2242 is an HBL with remarkably strong gamma-ray emission, making it an interesting object for further studies at energies above $\sim 1 \rm \, MeV$.

The interesting region where the inverse Compton branch becomes dominant
over the synchrotron branch lies 
in the energy range covered by the {\it INTEGRAL} instruments ($20 - 8000 \keV$).
Given the X-ray spectrum shown by RX J1211+2242 in the {\it BeppoSAX} data, it will not be easily detectable by this mission, even when assuming that the power law with $\alpha_X \simeq 1.0$ extends up to several hundred keV.
We have performed simulations that show that a 500 ksec observation would give a $< 5 \sigma$ detection by the SPI spectrometer.
However, the upcoming gamma-ray missions {\it GLAST} \cite{GLAST} and {\it AGILE} \cite{AGILE}, with their improved sensitivity, should be able to detect
 RX J1211+2242 in the EGRET band.
 
Even though this source is likely to be a strong TeV emitter, it will not be seen by most Cherenkov telescopes as the high energy emission of an object at redshift $z = 0.455$ will be suppressed by interaction of the TeV photons with the extragalactic infrared background \cite{kneiske}.
Nevertheless the {\it MAGIC} telescope \cite{MAGIC} with its low energy threshold of $\sim 30 \, \rm GeV$ is suitable to observe the peak of the IC component making RX J1211+2242 one of the most promising targets in the gamma-ray sky.

\begin{acknowledgements}
This research has made use of the NASA/IPAC Extragalactic Database (NED) which is operated by the Jet Propulsion Laboratory and of the SIMBAD Astronomical Database which is operated by the Centre de Donn\'ees astronomiques de Strasbourg. We thank Laura Maraschi for fruitful discussions and Steve Sturner for proof reading the manuscript. This work has received partial financial support from the Italian Space Agency and from COFIN grant 2001 028773-007.
\end{acknowledgements}

%
%
\clearpage
%
%
\clearpage
\begin{table*}
\caption[]{Journal of \sax observations}
\begin{tabular}{lrrrrr}
\tableline\tableline
obs. date & LECS & LECS & MECS & MECS\\
& exp. time & net counts & exp. time & net counts\\
\hline
 12/27/99 & 19191 s& $629.8 \pm 31.8$ & 46421 s& $1786.6 \pm 46.3$\\
 12/28/01 & 20268 s& $841.2 \pm 47.1$ & 56607 s& $3449.1 \pm 93.1$\\
 01/11/02 &  5011 s& $163.5 \pm 30.9$ & 16356 s&  $811.0 \pm 45.6$\\
\tableline
\end{tabular}
\label{journal}
\end{table*}
\begin{table*}
\caption[]{\sax: Best fit results for a single-power law model with free $N_H$}
\begin{tabular}{llllcrl}
\tableline\tableline
Observation & $\alpha_X$ & $N_H^{a}$ & $f_X^{b}$ & $f_X^{c}$& Nm$^{d}$ & $\chi_{\nu}^2 (dof)$\\
 &  &(Fit) &      &       &  &\\  
\hline
December 99  & 1.08 $+0.10 \atop -0.10$ & 3.49 $+1.24 \atop -0.87$& 2.30  & 1.90 & 0.79 & 0.89 (67)\\ 
December 01  & 0.94 $+0.05 \atop -0.05$ & 2.83 $+0.57 \atop -0.47$& 7.07  & 5.13 & 0.67 & 1.05 (145)\\ 
January  02 & 1.07 $+0.11 \atop -0.11$ & 3.22 $+1.61\atop -1.04 $& 5.15  & 4.50 & 0.62 & 0.97 (42)\\ 
Jan02 + Dec01& 0.95 $+0.03 \atop -0.04$ & 3.09 $+0.66\atop -0.53 $& 6.66  & 4.83 & 0.70 & 1.19 (170)\\
\tableline
\end{tabular}
\label{powfreesax}

$^{a}$ hydrogen column density in $\rm \times 10^{20} cm^{-2}$\\
$^{b}$ un-absorbed flux in $10^{-12}\; \rm erg\,cm^{-2}\,s^{-1}$ in
the 2--10 keV MECS energy band\\ 
$^{c}$ un-absorbed flux in $10^{-12}\;
\rm erg\,cm^{-2}\,s^{-1}$ in the 0.5--2.0 keV LECS energy band\\
$^{d}$ normalization of LECS versus MECS
\end{table*}
\begin{table*}
\caption[]{Journal of optical observations}
\begin{tabular}{lrrrrrrr}
\tableline\tableline
obs. date & B [mag] & error [mag]\\
\hline
 04/28/00  & 19.08 & 0.05 \\
 12/27/01A & 19.19 & 0.09 \\
 12/27/01B & 19.25 & 0.07 \\
 12/28/01A & 19.24 & 0.10 \\
 12/29/01B & 19.06 & 0.09 \\
\tableline
\end{tabular}
\label{journalobs}
\end{table*}
\end{document}